\documentclass[prd,nofootinbib]{revtex4-1}
\pdfoutput=1 % if your are submitting a pdflatex (i.e. if you have
\usepackage{dcolumn}% Align table columns on decimal point
\usepackage{graphicx}
\usepackage{color}
\usepackage{amsmath}
\usepackage{amssymb}
\usepackage{amsthm}
\usepackage{latexsym}
\usepackage{bm}
\usepackage{multirow}
\usepackage{cancel}
\usepackage{hyperref}
\usepackage{tabularx}
\newcommand*{\gev}{{\rm GeV}}
\newcommand*{\tev}{{\rm TeV}}

\newcommand*{\beq}{\begin{equation}}
\newcommand*{\eeq}{\end{equation}}
\newcommand*{\bea}{\begin{eqnarray}}
\newcommand*{\eea}{\end{eqnarray}}

\def\A{\ensuremath{A}}

\newcommand*{\mh}{\ensuremath{m_h}}
\newcommand*{\mA}{\ensuremath{m_A}}

\newcommand*{\hpm}{\ensuremath{H^\pm}}
\newcommand*{\hob}{\ensuremath{H_{\rm obs}}}

\newcommand*{\mhpm}{\ensuremath{m_{H^\pm}}}

\newcommand*{\half}{\ensuremath{\frac{1}{2}}}

\begin{document}

\begin{flushright}
KIAS-P17035\\
\end{flushright}
\vspace*{2.0cm}
\title{Identifying a light charged Higgs boson at the LHC Run II$^\dagger$}
 
\author{Abdesslam Arhrib$^{1}$, Rachid Benbrik$^{2,3}$, Rikard Enberg$^{4}$, \\ William Klemm$^{4,5}$, Stefano Moretti$^{6}$ and Shoaib Munir$^{7}$} 

\affiliation{$^1$ Facult\'{e} des Sciences et Techniques, Abdelmalek Essaadi University, B.P. 416, Tangier, Morocco \\
$^2$ LPHEA, Semlalia, Cadi Ayyad University, Marrakech, Morocco \\
$^3$ MSISM Team, Facult\'{e} Polydisciplinaire de Safi, Sidi Bouzid, B.P. 4162, Safi, Morocco \\
$^4$ Department of Physics and Astronomy, Uppsala University, 
Box 516, SE-751 20 Uppsala, Sweden \\
$^5$ School of Physics \& Astronomy, University of Manchester, 
Manchester M13 9PL, UK\\
$^6$ School of Physics \& Astronomy,
University of Southampton, Southampton SO17 1BJ, UK \\
$^7$ School of Physics, Korea Institute for Advanced Study,
Seoul 130-722, Republic of Korea}

\date{\today}

\begin{abstract}

We analyse the phenomenological implications of a light Higgs boson, $h$, within the CP-conserving 2-Higgs Doublet Model (2HDM) Type-I, for the detection prospects of the charged \hpm\ state at Run II of the  Large Hadron Collider (LHC), assuming $\sqrt{s}=13$\ \tev\ as energy and ${\cal O}(100~{\rm fb}^{-1})$ as luminosity. 
When sufficiently light, this $h$ state can open up the bosonic decay channel $\hpm \to
W^{\pm(*)}h$, which may have a branching ratio significantly exceeding those of the $\hpm \to \tau\nu$ and $\hpm \to cs$ channels.  We perform a broad scan of the 2HDM Type-I parameter space, assuming the heavier of the two CP-even Higgs bosons, $H$, to be the observed SM-like state with a mass near 125 GeV. 
Through these scans we highlight regions in which $\mhpm < m_t +m_b$ that are still consistent with the most recent limits from experimental searches. We find in these regions that, when the $\hpm \to
W^{\pm(*)}h$ decay mode is the dominant one, the $h$ can be highly fermiophobic, with a
considerably large decay rate in the $\gamma\gamma$ channel. This can result in the total cross section of the $\sigma(pp\to \hpm h \to W^{\pm(*)} + 4\gamma)$ process reaching up to ${\cal O}(100~{\rm fb})$. We therefore investigate the possibility of observing this spectacular signal at the LHC Run II.

\vspace*{4.0cm}
\centerline{\sl $^\dagger$We dedicate this work to the late Professor Maria Krawczyk, a friend and inspiration to us all.}

\end{abstract}

\preprint{KIAS-}

\maketitle
\section{Introduction}

The discovery of a resonance around 125\ \gev\ at the Large Hadron Collider
(LHC)~\cite{Aad:2012tfa,Chatrchyan:2012xdj} triggered plenty of
activity in the particle physics community. Comprehensive analyses to 
investigate the spin and parity of the discovered particle have confirmed
 its scalar nature. The measured signal rates of this scalar particle, \hob, in 
its dominant decay channels, agree with those predicted for the SM Higgs boson 
at the $2\sigma$ level~\cite{Khachatryan:2016vau}. 
However, the possibility that the \hob\ could belong to a model with an 
extended Higgs sector, such as the SM with an extra singlet, doublet and/or triplet has not been ruled out. 
Amongst such higher Higgs representations, those with an extra doublet or triplet 
also contain one or more charged Higgs bosons in their scalar spectrum. 
The discovery of such charged Higgs bosons would be an eminent 
signal of an extended Higgs sector and clear evidence of physics 
Beyond the SM (BSM).  

Among the simplest extensions of SM is the 2-Higgs Doublet Model (2HDM) 
in which the SM, containing a complex scalar doublet, $\phi_1$, is augmented by
 another doublet, $\phi_2$,
in order to give masses to all the fermions and gauge bosons. 
After Electro-Weak Symmetry Breaking (EWSB), 
out of the 8 degrees of freedom of the two Higgs doublets, 3 are eaten up by the EW gauge bosons to make up their
 longitudinal components, while the remaining 5 should manifest themselves as physical particles.
Therefore, the CP-conserving Higgs sector of the 2HDM contains 
three neutral Higgs bosons, two scalars ($h$ and $H$, with $m_h<m_H$),
 a pseudoscalar ($A$), and an \hpm\ pair. The requirement 
that one out of $h$ and $H$ 
have properties consistent with the \hob\ puts rather 
stringent bounds on the 2HDM parameter space. 
It is well known that, in a 2HDM, there exists a `decoupling limit', 
where $m_{H,A,H^\pm}\gg m_Z$~\cite{Gunion:2002zf}, and the couplings 
of the $h$ to the SM
particles are identical to those of the SM Higgs boson. 
Alternatively, the model also possesses an `alignment
limit', in which either one of
$h$~\cite{Carena:2013ooa,Bernon:2015qea} or $H$~\cite{Ferreira:2012my,Bernon:2015wef} can mimic the SM Higgs boson. 

The masses of the other two neutral Higgs bosons as well as the \hpm\ are also strongly
constrained by the results from their direct searches at various collider
experiments. Moreover, indirect constraints on these come from
 $B$-physics and precision EW measurements. In
 general, when the \hpm\ state is lighter than the sum
 of the masses of the top and bottom quarks, its dominant decay mode is
 $\tau\nu$. The ATLAS and CMS collaborations have released exclusion limits on
 the Branching Ratio (BR) of a generic \hpm\ state in this decay
 mode~\cite{Aad:2014kga,Khachatryan:2015qxa}. The only other decay channel of
 the \hpm\ probed at the LHC thus far is $c{s}$, but the resulting
 limits~\cite{Aad:2013hla,Khachatryan:2015uua} are rather weak compared to
 those for the $\tau \nu$ channel. In fact, in a specific (so called `flipped') 2HDM, the \hpm\ $\to$ $cb$
decay can also be relevant \cite{Akeroyd:2012yg} and has indeed been searched for by CMS \cite{CMS-PAS-HIG-16-030}.

However, in a recent study~\cite{Arhrib:2016wpw}, it was shown that in another specific 2HDM (called Type-I, henceforth 2HDM-I) with a SM-like $h$ and $\mhpm < m_t +m_b$, instead of these conventional channels, $W^{\pm (*)}h$ and/or $W^{\pm (*)}A$ can alternatively become the dominant decay
 channels of the \hpm, thereby competing with the fermionic modes \cite{Akeroyd:2016ymd,Moretti:2016qcc,Krawczyk:2017sug}. In another study~\cite{Enberg:2016ygw} it was shown that, when instead the $H$ is SM-like, it is possible for $h$ and $A$ to have masses such that 
 $m_h+m_A< m_Z$, without being in conflict with the direct search
 limits. Two other important features of the relevant parameter space were also noted there: i) for consistency with the EW
 precision measurements, such light $h$ and $A$ are
 accompanied by a \hpm\ not much heavier than the $Z$ boson, and ii) the $h$ can be extremely fermiophobic, so that its decays into SM fermions are highly suppressed, which can in turn result in a very large BR($h \to \gamma\gamma$).
In this study, we further explore this possibility of a light \hpm\ in the
 2HDM-I decaying via such a fermiophobic $h$ and the $W^{\pm (*)}$, along the lines of~\cite{Akeroyd:1998dt}. When
 the $\hpm$ is produced in the process $q\bar q' \to W^{\pm (*)} \to \hpm h$,
 due to the additional pair of photons coming from the second $h$, a very
 clean $W^{\pm (*)} + 4\gamma$ signal may result.

The paper is organised as follows. In Sect.~\ref{sec:2hdm} we briefly
review the various types of the 2HDM. In Sect.~\ref{sec:fphobic} we 
discuss parameter space regions, satisfying the
theoretical and experimental constraints available, where  
a light $h$, accompanied by a light \hpm\ and a 
SM-like $H$, can be obtained, while in 
Sect.~\ref{sec:w4gamma} we analyse the 
$W^{\pm}+4\gamma$ signal.\footnote{For the remainder of this letter, we suppress the ``$(*)$'' superscript from any $W^{\pm (*)}$ resulting from the decay of a charged Higgs.  An off-shell $W^{\pm*}$ is implied whenever $m_{\hpm}<m_h+m_W$.} In  Sect.~\ref{sec:discovery} we then discuss the potential visibility of the 
$\hpm$ in this final state at the 13\ \tev\ LHC.
We present our conclusions in Sect.~\ref{sec:conc}.

\section{\label{sec:2hdm} Types of the 2HDM}
The most general 2HDM scalar potential which is both 
$SU(2)_L\otimes U(1)_Y$ and CP invariant  is written as
\bea \label{eq:potential}
V(\phi_1,\phi_2)  &=& m_{11}^2\phi_1^\dagger\phi_1+ m_{22}^2\phi_2^\dagger\phi_2
-[m_{12}^2\phi_1^\dagger\phi_2+ \, \text{h.c.} ] \nonumber \\
&+&\half\lambda_1(\phi_1^\dagger\phi_1)^2
+\half\lambda_2(\phi_2^\dagger\phi_2)^2
+\lambda_3(\phi_1^\dagger\phi_1)(\phi_2^\dagger\phi_2) \nonumber\\ 
&+&\lambda_4(\phi_1^\dagger\phi_2)(\phi_2^\dagger\phi_1)
+ [\half\lambda_5(\phi_1^\dagger\phi_2)^2 + 
\text{h.c.}], 
\eea
where $\phi_1$ and $\phi_2$ have weak hypercharge $Y=+1$, while $v_1$ and
$v_2$ are their respective Vacuum Expectation Values (VEVs). 
Through the minimisation conditions of the potential, $m_{11}^2$ and $m_{22}^2$ 
can be traded for $v_1$ and $v_2$
and the tree-level mass relations allow the quartic couplings 
$\lambda_{1-5}$ to be substituted by the four 
physical Higgs boson masses and the neutral sector mixing term 
 $\sin(\beta-\alpha)$, where $\beta$ is defined through 
 $\tan\beta=v_2/v_1$, and $\alpha$ is the mixing angle between the 
CP-even interaction states. Thus, in total, the Higgs sector of the 2HDM 
 has 7 independent parameters, which include $\tan\beta$,
 $\sin(\beta-\alpha)$, $m_{12}^2$ and the four physical Higgs boson masses.

If both the Higgs doublets of a 2HDM couple to all fermions,  
they can mediate Flavor Changing Neutral Currents (FCNCs) at the tree level. 
In order to avoid large FCNCs, a $Z_2$ symmetry may be imposed
such that each type of fermion only couples to one of the
doublets~\cite{Glashow:1976nt}. The potential in Eq.~(\ref{eq:potential}) is thus
invariant under the symmetry $\phi_1 \to -\phi_1$ 
up to the soft breaking term proportional to $m_{12}^2$. 
Depending on the $Z_2$ charge assignment of the Higgs doublets, there are  
four basic Types of 2HDMs~\cite{Branco:2011iw,Gunion:2002zf}. 
In the Type-I model, only the doublet $\phi_2$ couples to all 
the fermions as in the SM. In the Type-X (or IV or ‘lepton-specific’) model, 
the charged leptons couple to $\phi_1$ while all the quarks couple to
$\phi_2$. In the Type-II model $\phi_2$ couples to up-type quarks and $\phi_1$
to down-type quarks and charged leptons. Finally, in the Type-Y (or III or
‘flipped’) model $\phi_2$ couples to up-type quarks and leptons and
$\phi_1$ to down-type quarks.
Note that for $\sin(\beta-\alpha)\approx 1$, $h$ has couplings consistent with the SM Higgs boson, while $H$ is the SM-like Higgs boson for
 $\sin(\beta-\alpha)\approx 0$. 

The Yukawa interactions in terms of the neutral and charged 
Higgs mass eigenstates in a general 2HDM can be written as 
\begin{eqnarray}
-{\mathcal L}_\text{Yukawa}^\text{2HDM} &=& \nonumber
 \sum_{f=u,d,\ell} \frac{m_f}{v} \left(
\xi_f^{h} {\overline f}f h
+ \xi_f^{H}{\overline f}f H
- i \xi_f^{A} {\overline f}\gamma_5f A
\right)    \\\nonumber & +&
\bigg\{\frac{\sqrt2V_{ud}}{v}\,
\overline{u} \left( m_u \xi_u^{A} \text{P}_L
+ m_d \xi_d^{A} \text{P}_R \right)
d H^{+} \\ &+&
\frac{\sqrt2m_\ell\xi_\ell^{A}}{v}\,
\overline{\nu_L^{}}\ell_R^{}H^+
+\text{h.c}\bigg\},    
\label{Eq:Yukawa}
\end{eqnarray}
where $v^2=v_1^2+v_2^2=(2\sqrt{2} G_F)^{-1}$, $V_{ud}$ is the top-left
entry of the Cabibbo-Kobayashi-Maskawa
 (CKM) matrix, and $P_L$ and $P_R$ are the 
left- and right-handed projection operators, respectively.  
In the 2HDM-I, $\xi^{h}_f=\cos\alpha/ \sin\beta$ and $\xi^{H}_f=\sin\alpha/
\sin\beta$, for $f=u, d, l$, while $\xi^{A}_d=- \cot\beta$,
$\xi^{A}_u=\cot\beta$, and $\xi^{A}_l=- \cot\beta$.

As pointed out earlier, experimental searches can tightly constrain the properties of the \hpm\ in a 2HDM, depending on its Type.
For instance, in the Type-II and Type-Y 2HDMs, the measurement of 
the BR($b\to s\gamma$) constrains \mhpm\ to be larger than about 570\
\gev~\cite{Misiak:2017bgg,Misiak:2015xwa}, which makes these models irrelevant for this study. We therefore focus here on the 2HDM-I, in which one can still 
obtain a \hpm\ with a mass as low as $\sim
100-200$\ \gev~\cite{Misiak:2017bgg,Enomoto:2015wbn,Hussain:2017tdf}, provided that $\tan\beta\geq 2$.  
 
%%%%%%%%%%%%%%%%%%%%%%%%%%%%%%%%%%%%%%%%%%%%%%%%%%%%%%%%%%%%
\section{\label{sec:properties} Production via $pp\to H^\pm h$ and decay through $H^\pm \to W^\pm h$}
%%%%%%%%%%%%%%%%%%%%%%%%%%%%%%%%%%%%%%%%%%%%%%%%%%%%%%%%%%%%
In our analysis, we concentrate on the scenario where $H$ is the SM-like
Higgs, while $h$ is lighter than 125 GeV and $A$ could be either lighter
or heavier than the $h$. In this section we first discuss the light CP-even Higgs $h$ in our scenario, which can occur near the alignment limit ($\sin(\beta-\alpha) \approx 0$)~\cite{Enberg:2016ygw}, and show how it can be highly fermiophobic, decaying dominantly to two photons. We then proceed to 
$pp\to W^{\pm*}\to H^\pm
h$ production via $s$-channel $W^\pm$ exchange,
followed by the $H^\pm \to W^\pm h$ decay mode, which
could be the dominant one, allowing the \hpm\ to escape the Tevatron and
LHC limits, which are based on the fermionic decay modes, $H^\pm\to \tau\nu, cs,
cb$ \cite{Khachatryan:2015qxa,Aad:2014kga,CMS-PAS-HIG-16-030}.

%%%%%%%%%%%%%%%%%%%%
\subsection{\label{sec:fphobic} Fermiophobic $h$ in the 2HDM-I}
%%%%%%%%%%%%%%%%%%%%
It is well known that, in the SM, the $h\to \gamma\gamma$ decay is dominated by the $W^\pm$ loop, which is partly cancelled by a sub-leading contribution from the top quarks; in the 2HDM, we additionally have a \hpm\ contribution. The $W^\pm$ loop depends on $hW^+W^-\propto \sin(\beta-\alpha)$, the fermionic loops on $hf\bar{f}\propto\cos\alpha/\sin\beta$, and the \hpm\ contribution enters through the triple scalar coupling $hH^\pm H^\mp$, which depends on the scalar 
parameters of the potential. 

In our 2HDM-I scenario with $H$ being the SM-like Higgs boson,
the $W^\pm$ loops in $h\to \gamma\gamma$ 
get suppressed by this factor of $\sin(\beta-\alpha)\approx 0$. 
For the fermionic loops, $\cos\alpha$ is computed through
\begin{eqnarray}
\cos\alpha=\sin\beta  \sin(\beta-\alpha)+\cos\beta  \cos(\beta-\alpha).
\end{eqnarray}
For negative $\sin(\beta-\alpha)$ and positive $\cos(\beta-\alpha)$, it is clear that $\cos\alpha$ will vanish for a particular choice of $\tan\beta$. When this scenario takes place, since its couplings to fermions are proportional to $\cos\alpha$, the $h$ becomes fermiophobic~\cite{Akeroyd:1995hg}. 
Therefore, $h\to f\bar{f}$ and $h\to gg$ vanish. Moreover, since the $h$ of interest here is lighter than 120\,GeV, implying that the $h\to VV^*$ decay is phase-space suppressed, so the $h\to \gamma\gamma$ decay channel is expected to dominate in this limit. 

\begin{table}[tbp]
\begin{center}
	\begin{tabular}{cc}
\hline
Parameter & Scanned range \\
\hline 
\hline
		\mh\ (\gev) & (10, 120) \\
%		\mH\ (\gev) & 125 & 125 \\
		\mA\ (\gev) & (10, 500) \\
		\mhpm (\gev) & (80, 170) \\
		$\sin(\beta-\alpha)$  & ($-1$, 1) \\
$m_{12}^2$ (\gev$^2$) & (0, $m_{\A}^2\sin\beta\cos\beta$) \\
		$\tan\beta$ & (2, 25) \\
\hline
	\end{tabular}
	\caption{Scanned ranges of the 2HDM-I parameters.}
	\label{tab:params}
\end{center}
\end{table}

\begin{figure}[tbp]
\includegraphics[width=0.49\textwidth]{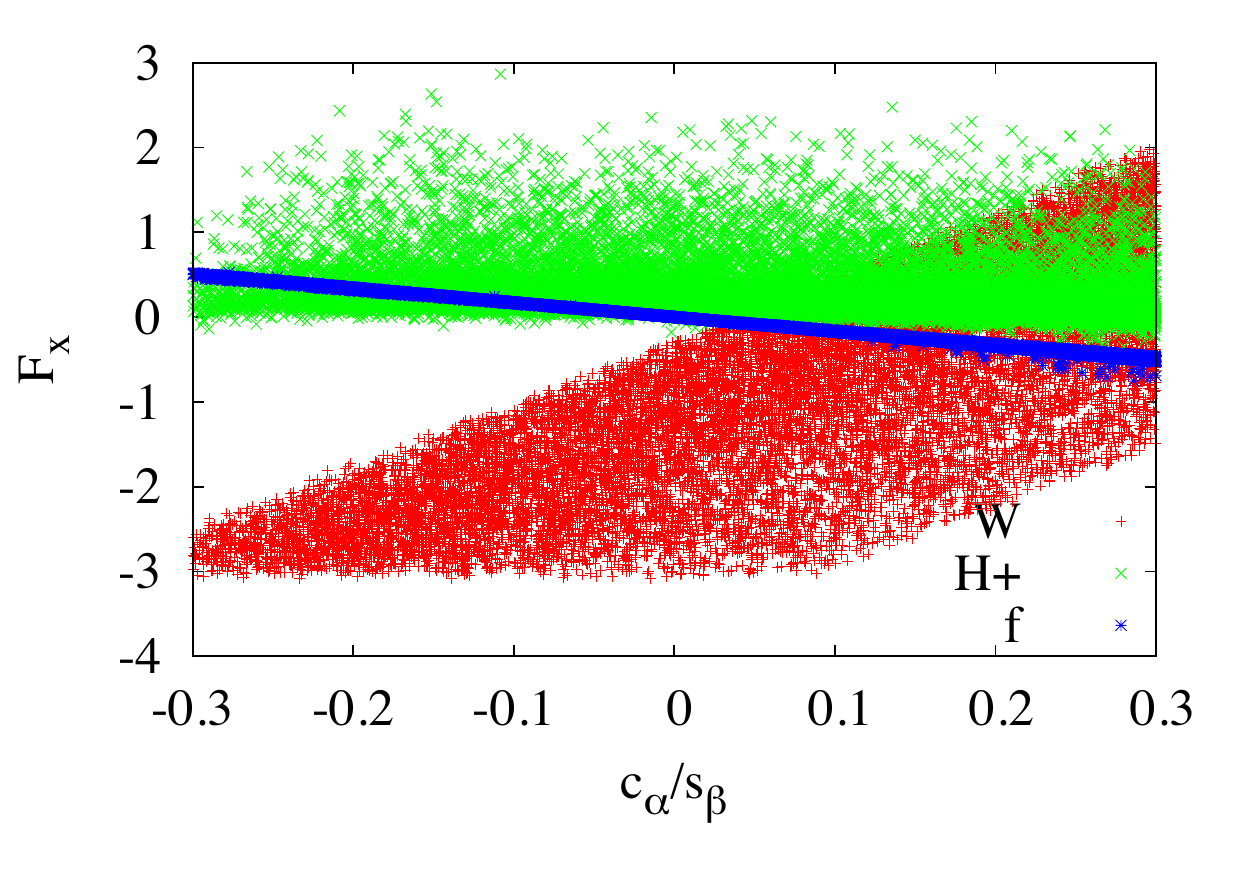}\includegraphics[width=0.49\textwidth]{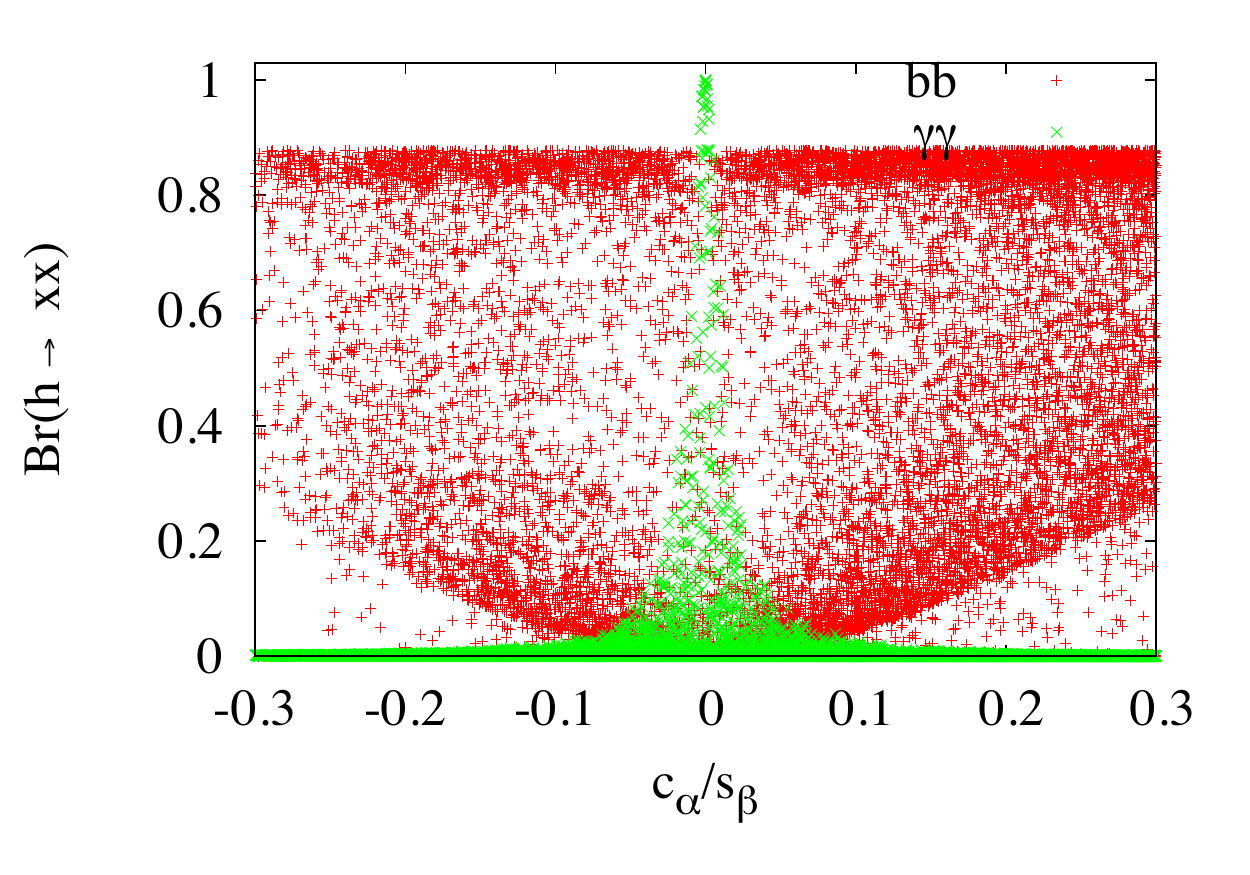}
\caption{Left: Contribution, $F_x$,  defined in the text, to the $h\to \gamma\gamma$ decay, corresponding to the $W^\pm$ (red), fermions (blue), and $H^\pm$ (green) loops. Right: BRs of the $h\to  \gamma\gamma$ (green) and $h\to b\bar{b}$ (red) decays.}
\label{fig:hgg}
\end{figure}

To demonstrate this effect, we performed a systematic numerical scan of the 2HDM-I parameters over the ranges indicated in Tab. \ref{tab:params} (with $m_H$ fixed to 125\ \gev) using the 2HDMC-v1.7.0~\cite{Eriksson:2009ws} program. In the left panel of Fig.~\ref{fig:hgg} we show the loop factors, $F_x$, corresponding to 
$W^\pm$, fermions, and \hpm\ as functions of the reduced coupling
$hf\bar{f}=\cos\alpha/\sin\beta$ for the points obtained from our scan. These loop factors are defined as
 \begin{eqnarray}
F_f &=& \sum_{i} N_{f} Q^2_{f} \xi_f^{h} F_{1/2}(\tau_{f}),  \nonumber \\
F_{H^\pm} &=& g_{hH^{\pm}H^{\mp}}\frac{m^2_W}{m^2_{H^\pm}}F_{0}(\tau_{H^\pm}), \\
F_W&=&F_{1}(\tau_{W}) \sin(\beta-\alpha), \nonumber
\end{eqnarray}
where $\tau_x=m_h^2 /(4m_x^2)$ and the scalar functions $F_{0,1/2,1}$ can be found in, e.g., \cite{Djouadi:2005gj} (our convention uses the opposite sign from \cite{Djouadi:2005gj}).
It is clear from the figure that, in most of the
cases, the $W^\pm$ loop is dominant and interferes destructively 
with the \hpm\ and top-quark loops. 
In the exact alignment limit, where $\sin(\beta-\alpha)\to 0$, the $W^\pm$ loops 
vanish and only the \hpm\ and top loops contribute,
interfering destructively. Away from the exact alignment limit, for certain values of $\sin(\beta-\alpha)$ and $\tan\beta$, $\cos\alpha$ vanishes.
 Therefore, as intimated, $h$ becomes fermiophobic, and consequently, as the right panel of Fig.~\ref{fig:hgg} further illustrates, the BR$(h\to \gamma\gamma)$ can become 100\% for $\cos\alpha/\sin\beta =0$. 

Several searches for fermiophobic Higgs bosons have been performed by the LEP and Tevatron colliders, imposing stringent limits. 
At LEP-II, a fermiophobic Higgs boson was searched for through $e^+e^- \to Zh$, 
 where $h$ decays to 2 photons~\cite{Abdallah:2003xf,Rosca:2002me}, and a lower limit of order 100\,GeV was set on the mass of a SM-like $h$.
Tevatron also searched for a fermiophobic Higgs boson produced via Higgs-strahlung, $pp\to Vh$ ($V=W^\pm,Z$), as well as vector boson fusion, $qq\to q'q'h$, with similar results~\cite{Abazov:2008ac} to those obtained at LEP-II. In our 2HDM-I scenario, since the $VVh$ coupling is suppressed due to $\sin(\beta-\alpha)\approx 0$, these limits from LEP and Tevatron would apply only weakly. However, one can also produce such Higgs bosons in association with a CP-odd Higgs boson through $e^+e^-\to hA$, which depends on the coupling $ZhA \propto \cos(\beta-\alpha)$. 

The complementarity of these $hZ$ and $hA$ searches with $h\to\gamma\gamma$ allowed the DELPHI collaboration to place stringent limits on $m_h$ and $m_A$ in fermiophobic models~\cite{Abdallah:2003xf}. These constraints only apply to exactly fermiophobic models, whereas in this work we are most interested in the parameter space close to, but not necessarily at, the fermiophobic limit.  The combined LEP $hZ$ limits can readily be applied to models which are not at the fermiophobic limit, and these are tested with HiggsBounds~\cite{Bechtle:2013wla}, but a similar application of the DELPHI fermiophobic $hA$ results, which depend on $m_h$ and $m_A$, is less straightforward and not included in HiggsBounds. In Appendix \ref{app:halimit}, we describe a method for approximating the $hA$ limits more generally, which we apply to our scan in Sect.~\ref{sec:w4gamma}.  We note that the OPAL collaboration performed a similar search~\cite{Abbiendi:2002yc}, but their limits are weaker than the ones we apply here.

Following the work of Refs.~\cite{Akeroyd:2003bt,Akeroyd:2005pr}, 
the CDF collaboration has also searched for fermiophobic Higgs bosons~\cite{Aaltonen:2016fnw} in the $W^\pm+4\gamma$ channel highlighted in this paper.  This search should in principle have sensitivity to some of the parameter space closest to the fermiophobic limit. However, the CDF limits are presented only for the exactly fermiophobic scenario and are not readily extendable to our more general search.

As for the LHC, despite the fact that a phenomenological framework for a 4-photon search was set up in~\cite{Akeroyd:2003xi} (also covering the $H^\pm h$ production mode addressed here), no ATLAS and CMS 
experimental analyses on these lines exist to date. One thus has to rely on Ref.~\cite{Delgado:2016arn}, which
uses data for the 2 photon Higgs search to constrain the scenario where 4 photons are produced. This study, however, does not exclude the region of parameter space discussed here, while the $H^\pm h\to W^{\pm}hh\to W^{\pm} + 4\gamma$ analysis of Ref.~\cite{Akeroyd:2003xi} captures a somewhat different region of parameter space from the one considered here, with $m_{H^\pm}>100$\,GeV and $m_h>40$\,GeV, and only considers the exact fermiophobic limit. Our present study 
extends to much lower masses of both of these Higgs bosons (down to $m_{H^\pm}\approx80$\,GeV and $m_h\approx$ 10\,GeV) and considers a less restrictive range of values for the other model parameters.\footnote{Furthermore, we perform here a detailed kinematical analysis of the 4-photon signal and background which was missing in~\cite{Akeroyd:2003xi}.}
 
\subsection{\label{sec:w4gamma} $pp\to H^\pm h$ and its $W^\pm +
  4\gamma$ final state}
We performed a numerical scan of the parameter space given in Tab.~\ref{tab:params} to investigate the scenarios which can result in a significant cross section for the $W^\pm +4\gamma$ final state.\footnote{The original scan, where all the input parameters were scanned uniformly, was supplemented by scans with $m_h<62.5$ GeV and $62.5 < m_h < 90$ GeV in order to obtain an appreciable density of points in each range. Apparent discontinuities in some figures at the resulting $m_h$ boundaries are a result of this choice.}
These points were checked for consistency with various experimental constraints
from direct Higgs searches, $B$-physics, and EW precision data. The complete list of the constraints imposed can be found in Sect.~2
of~\cite{Enberg:2016ygw}. We additionally required points to satisfy the constraints from fermiophobic $e^+e^-\to h A$ searches as described in Appendix~\ref{app:halimit}. 

The direct search constraints were checked using the latest stable version (v4.3.1) of the public code HiggsBounds~\cite{Bechtle:2013wla}. HiggsBounds~4 does not include searches from the 13 TeV LHC; we have therefore additionally checked our results against a beta version of HiggsBounds~5, which includes the recent searches.\footnote{T.\ Stefaniak, private communication (2017); see \url{http://higgsbounds.hepforge.org}.} We find that this rules out a small fraction of our parameter points but does not change the overall distribution very much, except for a slight tendency that more points at very low $\mhpm$ are ruled out.

\begin{figure}[h!]
\includegraphics[width=0.45\textwidth]{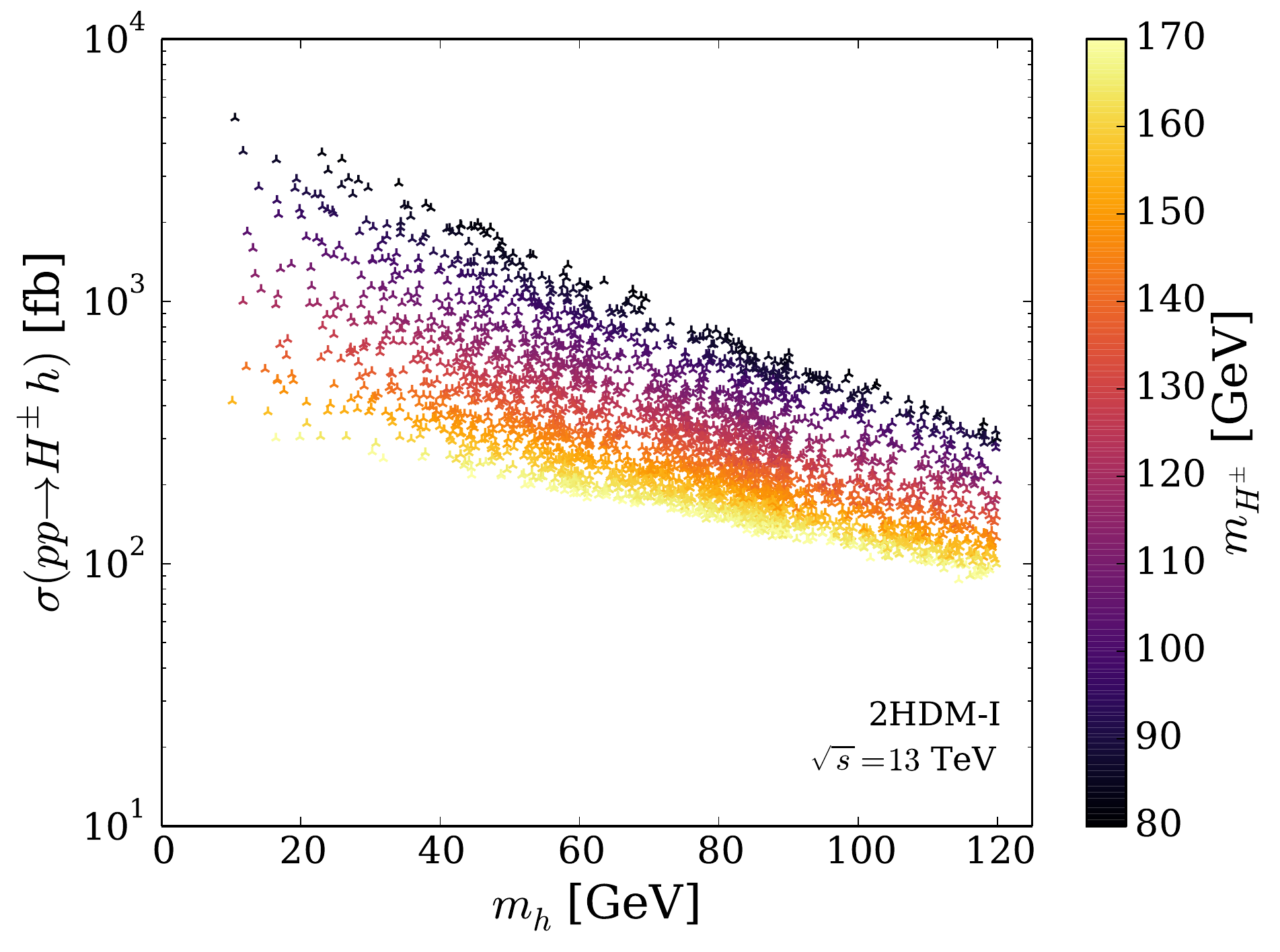}
\includegraphics[width=0.45\textwidth]{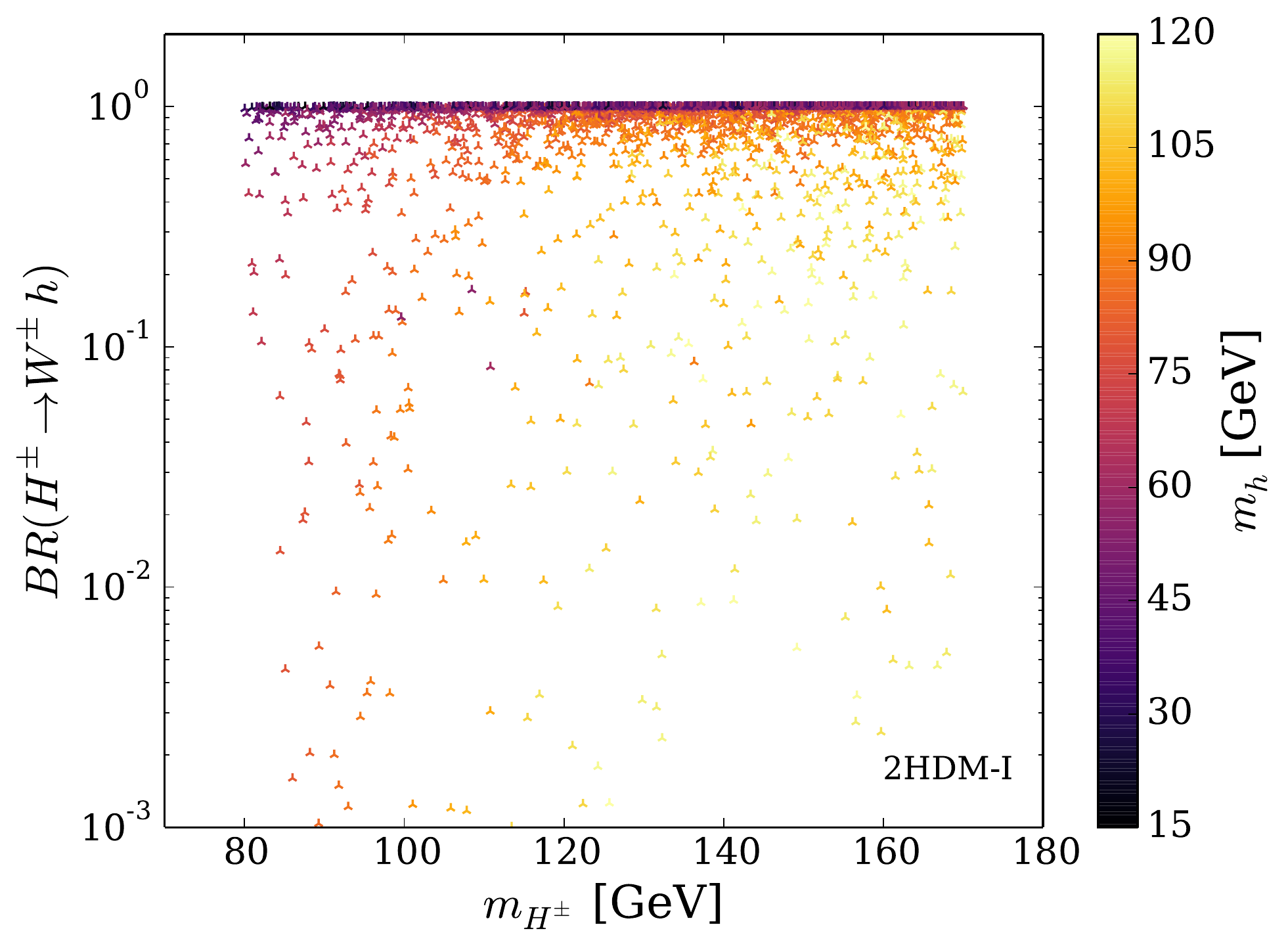}
\caption{Left: $\sigma(q q' \to H^\pm h)$ at $\sqrt{s}=13$~TeV 
 as a function of $m_h$ for the points passing the parameter space scan, with $m_{H^\pm}$ indicated by the colour map. Right: BR$(H^\pm\to W^\pm h)$ as a function of $m_{H^\pm}$, with $m_h$ indicated by the colour map.}
\label{fig:sigma-BR}
\end{figure}

For calculating the cross section for the process $qq'\to H^\pm h$
with $q=u,d,s,c,b$ (i.e., in the five-flavour scheme) at $\sqrt{s}=13~\tev$, 
we used 2HDMC combined with 
MadGraph5\_aMC@NLO\ \cite{Alwall:2014hca}. 
In the left panel of Fig.~\ref{fig:sigma-BR} we show the cross section $pp\to W^{\pm*} \to H^\pm h$ for the points obtained in our scan that pass all the constraints. The cross section has two sources of enhancement: the first is the 
$H^\pm W^\mp h$ coupling, which is proportional to $\cos(\beta-\alpha)$ 
and hence near-maximal in our scenario, while 
the second is the large phase space afforded due to a light $h$ and/or $H^\pm$.
It is clear that this production cross
section could reach the pb level for relatively light $h$, in the range 10--60\,GeV, 
 and light, 80--110 GeV, \hpm.  These cross sections can be comparable to, and in some cases exceed, the production of a light charged Higgs via top decay, e.g. $pp\to \bar{t}t\to \bar{t}bH^+$, especially at larger values of $\tan\beta$, where the coupling of $H^\pm$ to fermions is suppressed in 2HDM-I models.  Furthermore, the $\bar{t}bH^+$ channel does not give rise to the low-background $W^\pm+4\gamma$ signature considered here. 

\begin{figure}[h!]
\includegraphics[width=0.48\textwidth]{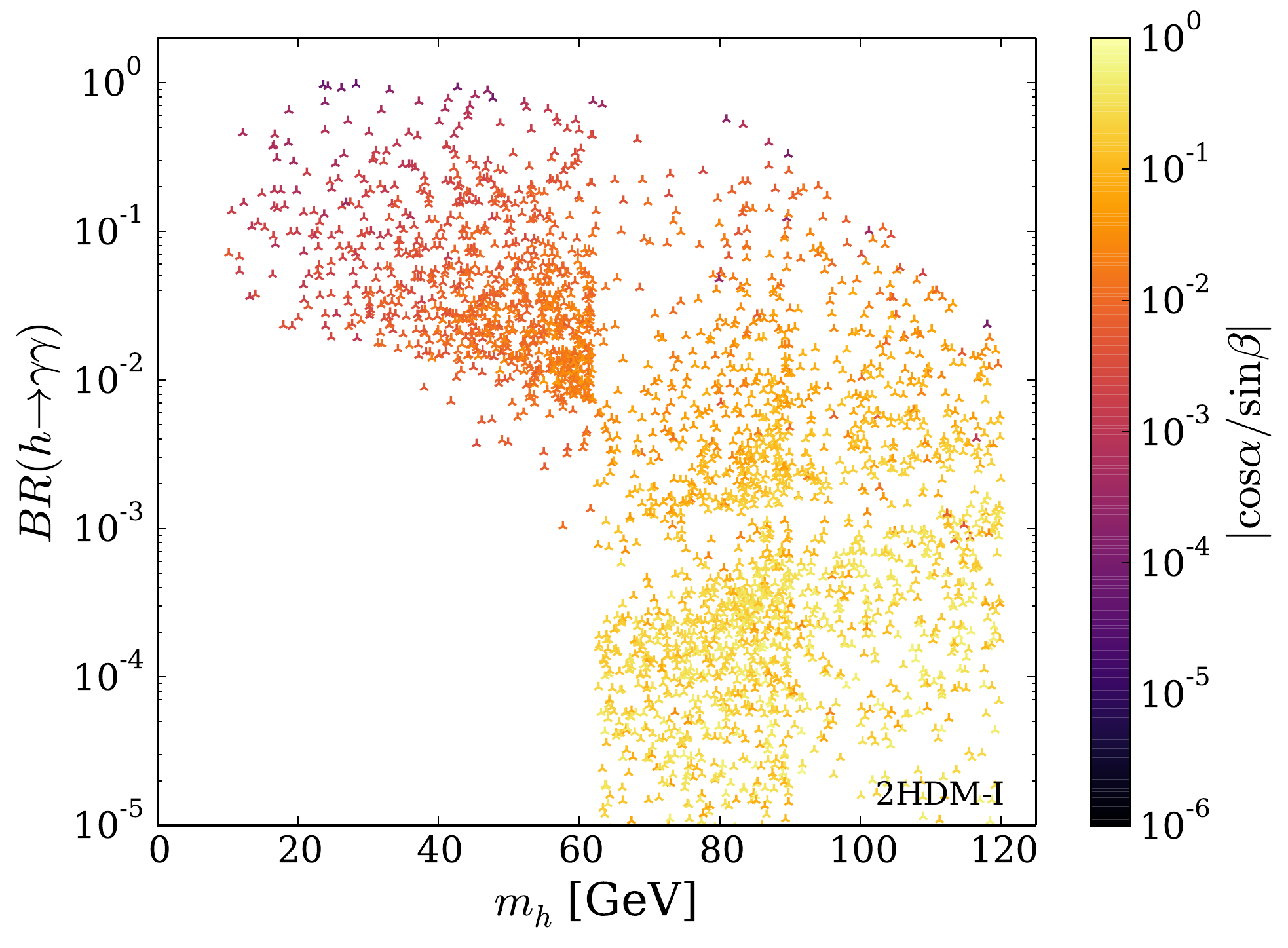}
\includegraphics[width=0.48\textwidth]{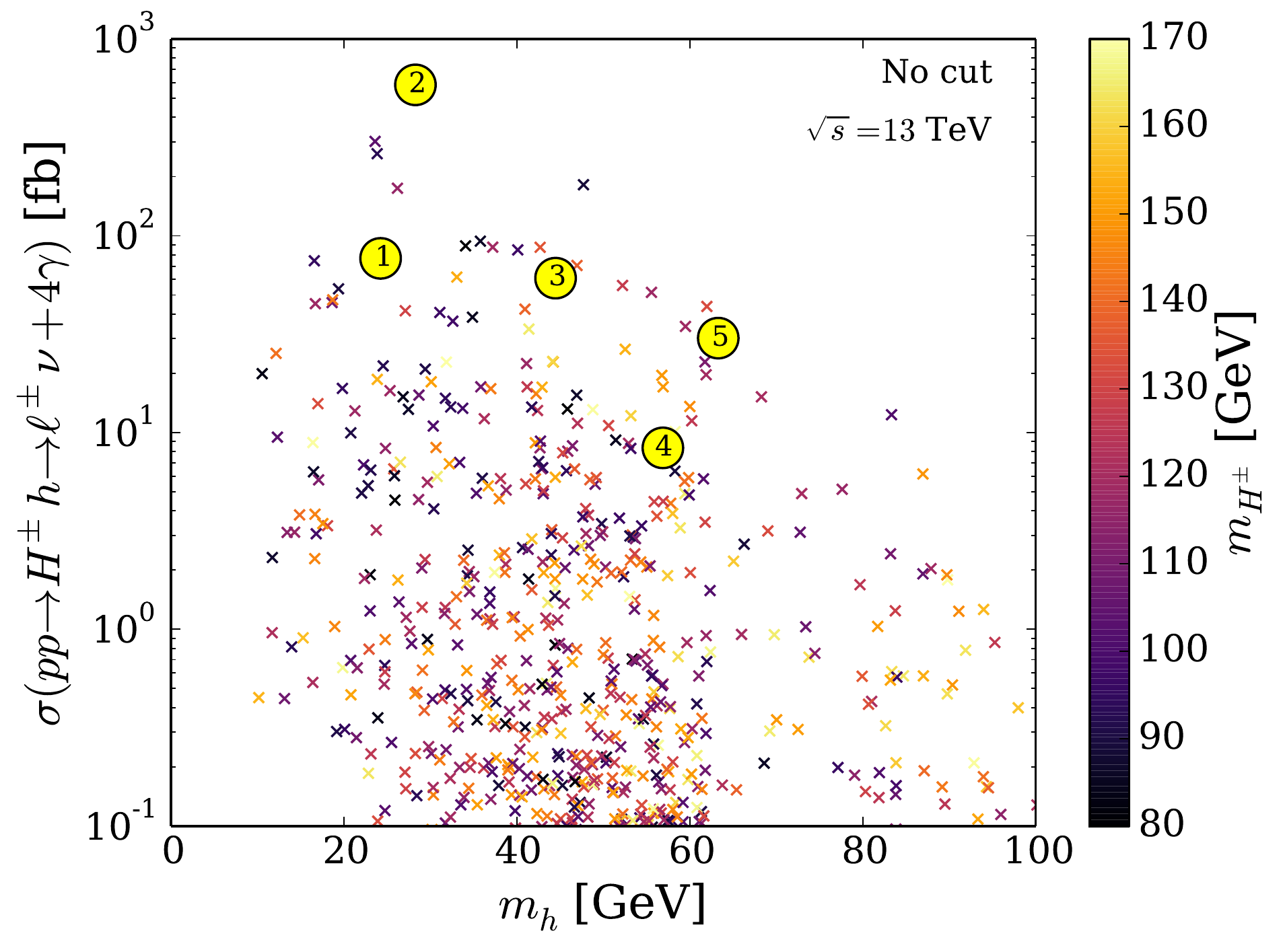}
\caption{Left: BR$(h\to\gamma\gamma)$ as a function of the mass of $h$, 
 with the heat map showing $|\cos\alpha/\sin\beta|$.
Right: Signal cross section, $\sigma(\ell^\pm\nu+4\gamma)$ as a function of $m_h$, with the heat map showing the mass of \hpm. The five selected benchmark points are highlighted in yellow circles.}

\label{fig:sigma-nocut}
\end{figure}

Similar to the $H^\pm h$ production, the decay $H^\pm \to W^{\pm} h$ 
 also enjoys the enhancement factor from $\cos(\beta-\alpha) \approx 1$.
The right panel of Fig.~\ref{fig:sigma-BR} illustrates that the
 BR$(H^\pm \to W^\pm h)$ can reach 100\% for a very light $h$. In the left panel of Fig.~\ref{fig:sigma-nocut} we show 
the BR($h\to \gamma\gamma$) as function of $m_h$,
 with the other 2HDM-I parameters varying in the ranges given in Table~\ref{tab:params}. We notice in the figure that before the opening of the $h\to WW^*$ channel, the BR($h\to \gamma\gamma$) could reach 100\% for small values of $\cos\alpha/\sin\beta$.
By putting together all these observations -- the large $H^\pm h$ cross sections, dominant $H^\pm\to W^\pm h$ decays, and the possibility of a fermiophobic $h$ that could decay primarily into two photons -- one can immediately anticipate a significant cross section for the $W^\pm hh\to W^\pm + 4\gamma$ final state.
This is confirmed by the right panel of Fig.~\ref{fig:sigma-nocut}, in which one sees that the total cross section for our signal, $\sigma(qq'\to H^\pm h \to W^\pm hh \to \ell^+ \nu + 4\gamma)$ (which we calculate as $\sigma(qq'\to H^\pm h)\times {\rm BR}(H^\pm \to W^\pm h)\times {\rm BR}(h\to \gamma\gamma)^2\times {\rm BR}(W^\pm\to\ell^\pm\nu)$), can reach the pb level for low $m_h$.

From these points, we have selected a few benchmark points (BPs), highlighted in the right panel of Fig.~\ref{fig:sigma-nocut}, which have a significant $W^\pm+4\gamma$ cross sections for varying values of $h$ and \hpm\ masses. The specifics of these BPs are given in Table~\ref{tab:BPs}.

\begin{table}[h]
\begin{center}
\begin{tabular}{c|c c c c c c|c }
BP & $m_h$ & $m_{H^\pm}$ & $m_A$ & $\sin(\beta-\alpha)$ & $m_{12}^2$ & $\tan\beta$ & $\sigma(W^\pm+4\gamma)$ [fb] \\\hline
1  & 24.2 &  152.2 &  111.1  & -0.048 &  19.0 &  20.9 &  359 \\
2  & 28.3 &  83.7 &  109.1  & -0.050 &  31.3 &  20.2 &  2740 \\
3  & 44.5 &  123.1 &  119.9  & -0.090 &  30.8 &  10.9 &  285 \\
4  & 56.9 &  97.0 &  120.3  & -0.174 &  243.9 &  5.9 &  39 \\
5  & 63.3 &  148.0 &  129.2  & -0.049 &  173.1 &  20.7 &  141
\end{tabular}
    \caption{Input parameters and parton-level cross sections (in fb) corresponding to the selected BPs. All masses are in GeV and for all points $m_H=125~\gev$. Here $\sigma(W^\pm+4\gamma) = \sigma(qq'\to H^\pm h)\times {\rm BR}(H^\pm \to W^\pm h)\times {\rm BR}(h\to \gamma\gamma)^2$ for the LHC at 13~TeV (in constrast to Figures \ref{fig:sigma-nocut} and \ref{fig:sigma-w-cuts}, a factor of $BR(W^\pm\to\ell^\pm\nu)$ is not included here).}
	\label{tab:BPs}
\end{center}
\end{table}

\section{\label{sec:discovery} Discovery potential}

\begin{figure}[h!]
\includegraphics[width=0.48\textwidth]{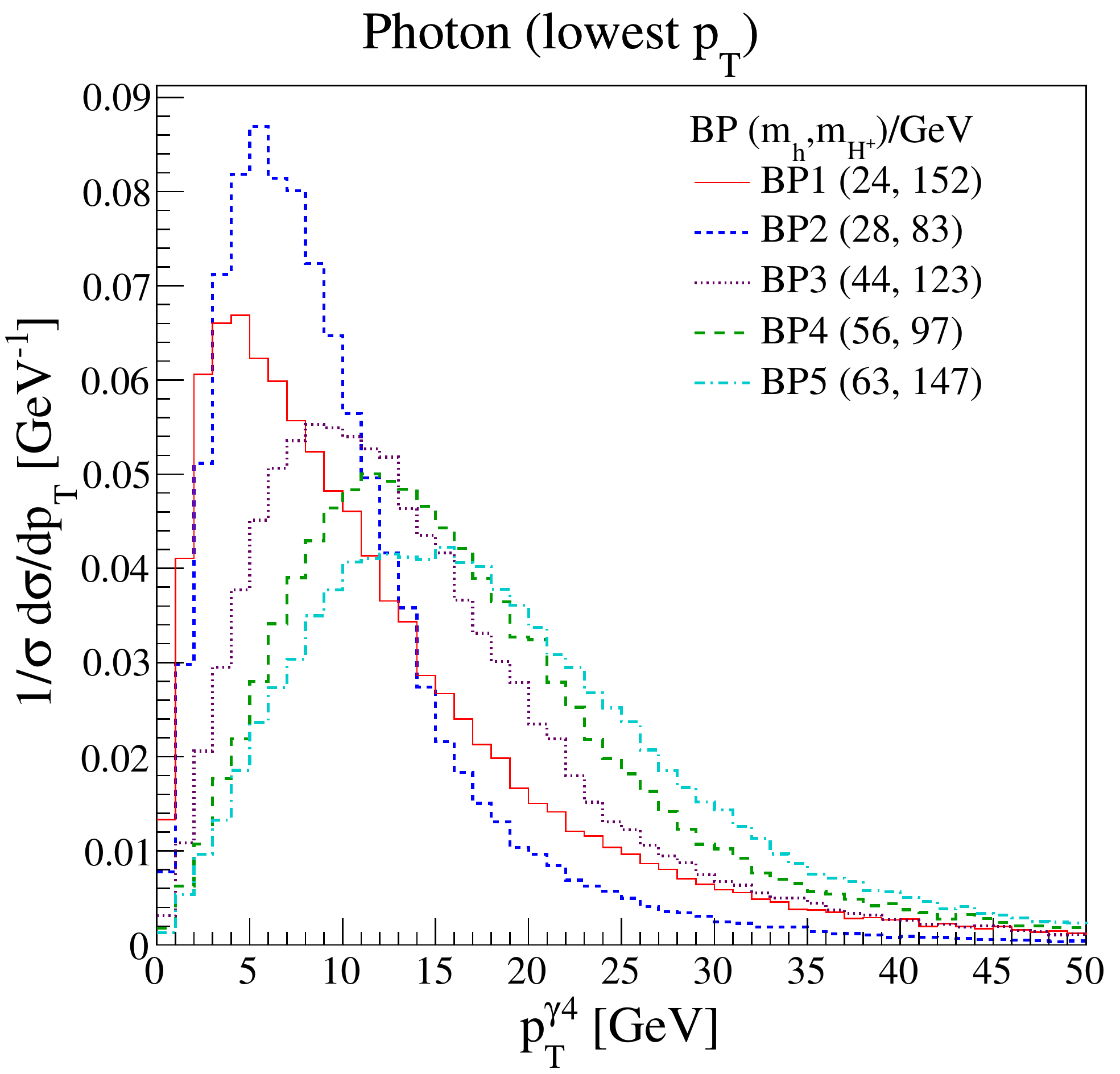}
\includegraphics[width=0.48\textwidth]{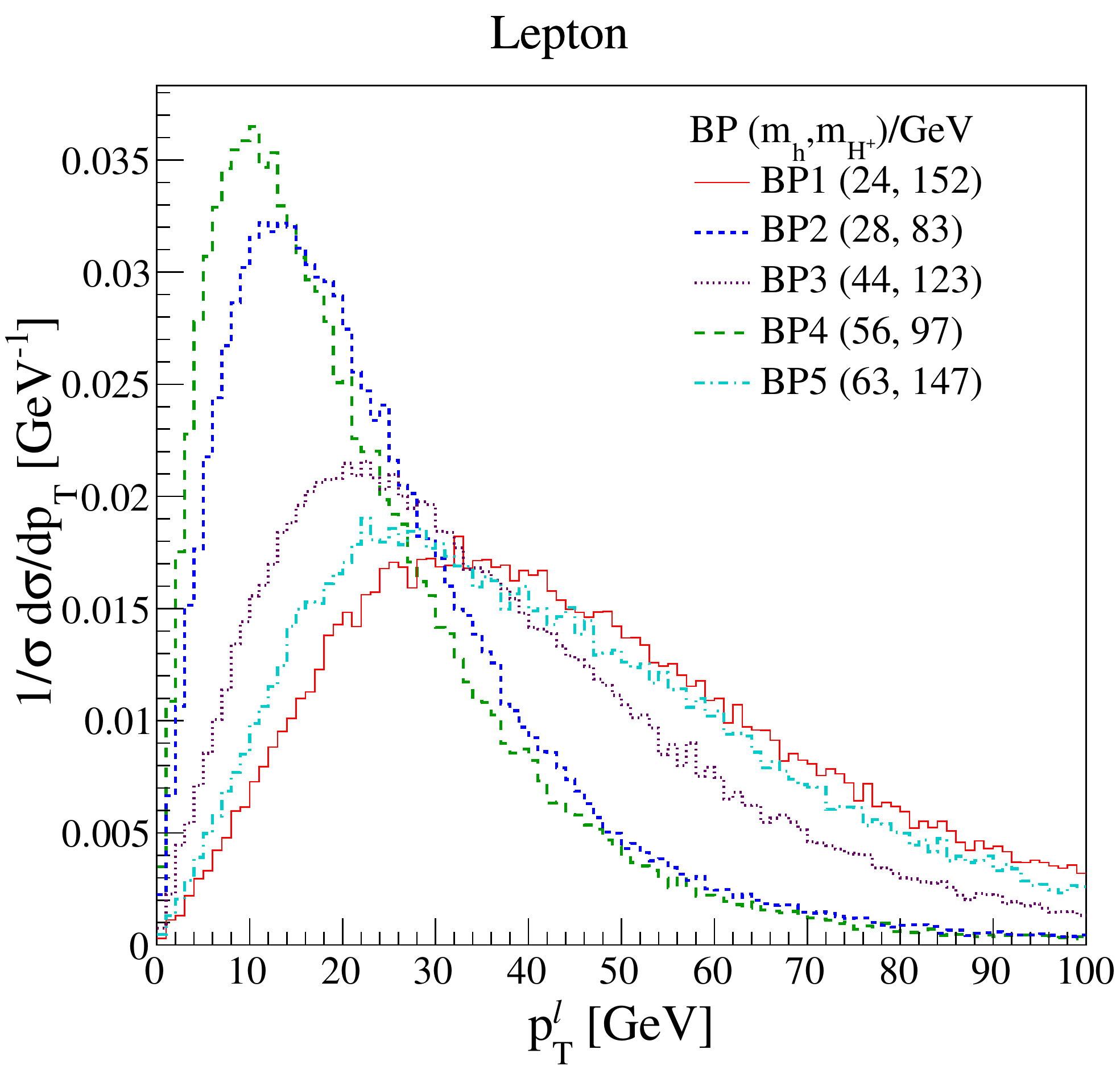}
\caption{Transverse momentum distributions for the softest photon (left) and the lepton (right) for the $\ell^\pm\nu + 4\gamma$ signal for the various BPs.}
\label{fig:kin-cuts}
\end{figure}

Next we consider the potential for the 13\ \tev\ LHC to observe this $W^\pm+4\gamma\to\ell^\pm\nu+4\gamma$
($\ell=e,\mu$) signature. Fig.~\ref{fig:kin-cuts} shows the distributions of the transverse momenta ($p_T$'s)
for one the BPs, for both the lepton and the softest photon. Both of them result from decays of relatively light intermediate states, so the distributions are skewed towards low $p_T$. The photon $p_T$, in particular, peaks at lower values for BPs with smaller $m_h$. The lepton $p_T$ distribution is sensitive to both $m_h$ and $m_{H^\pm}$, as is evident for the distribution corresponding to BP4, wherein the lepton coming from the off-shell $W^\pm$ tends toward low $p_T$, owing to the fact that $m_{H^\pm}-m_h$ is much smaller than $m_{W^\pm}$. Noting also that these distributions fall off rapidly in the $p_T$ ranges that might reasonably be used to select events, the discovery potential could be very sensitive to the choice of triggers and event selection criteria.

The experiments cannot trigger on such low-$p_T$ single photons or leptons, though, so the necessary triggers will have to be on combinations of multiple objects. For example, the ATLAS high level trigger (HLT) selection~\cite{Aaboud:2016leb,ATL-DAQ-PUB-2017-001} for a single isolated electron or muon goes down to 26\,GeV, with offline selection only slightly higher. Triggering on two muons, however, reduces the required momenta to 14\,GeV. Similarly, a single photon requires 120\,GeV in the HLT, but two tight photons require 22\,GeV each. It is therefore conceivable that the combinations required for the analysis we are proposing, for example, a lepton plus a photon trigger, or a four photon trigger, with low enough transverse momenta 10-15 GeV, could be added to the trigger menu.

Furthermore, it should be noted that the freedom of choice in selecting the optimal triggers is enabled by the fact that
the background for this process is essentially non-existent.
We estimated the irreducible SM $W^\pm+4\gamma$ background using MadGraph5\_aMC@NLO. Requiring, e.g., four photons and one lepton, all with $p_T>10$~GeV, along with pseudorapidity and isolation cuts described below, we find a cross section of less than $10^{-6}$~pb. In fact, we expect that instrumental backgrounds (e.g., mis-identification of a lepton or a jet as a photon~\cite{Aad:2009wy,GoyLopez:2016trw}), will not change this conclusion, as long as all 4 photons are indeed reconstructed.

\begin{figure}[h!]
\includegraphics[width=0.48\textwidth]{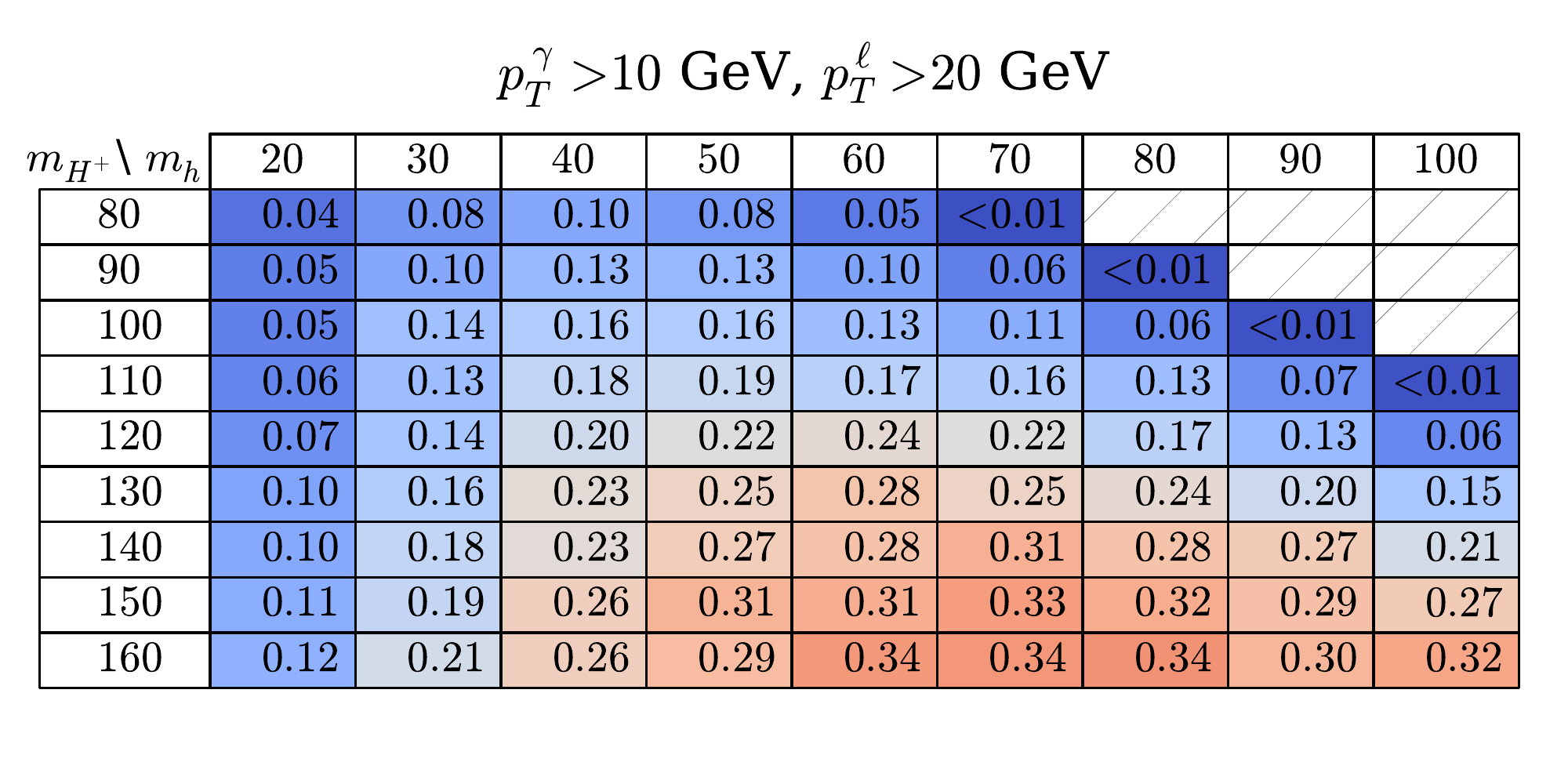}%
\includegraphics[width=0.48\textwidth]{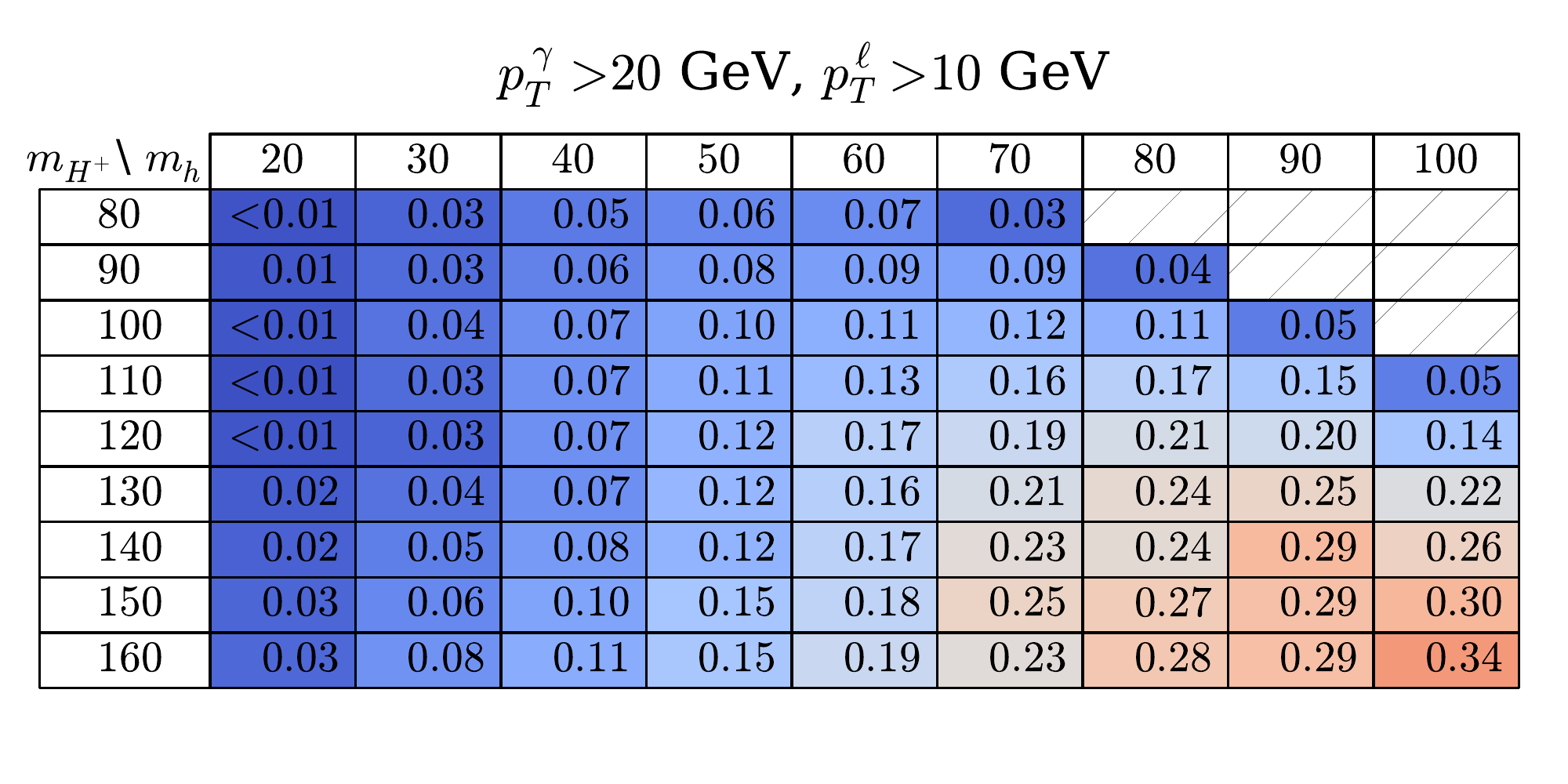}%
\caption{Efficiency, $\epsilon=\sigma(\text{cuts})/\sigma(\text{no cuts})$, for the $\ell^\pm\nu + 4\gamma$ final state for the two choices (i) and (ii) of cuts described in the text (left and right, respectively). All masses are in GeV.}
\label{fig:efficiencies} 
\end{figure}

\begin{figure}[h!]
\includegraphics[width=0.48\textwidth]{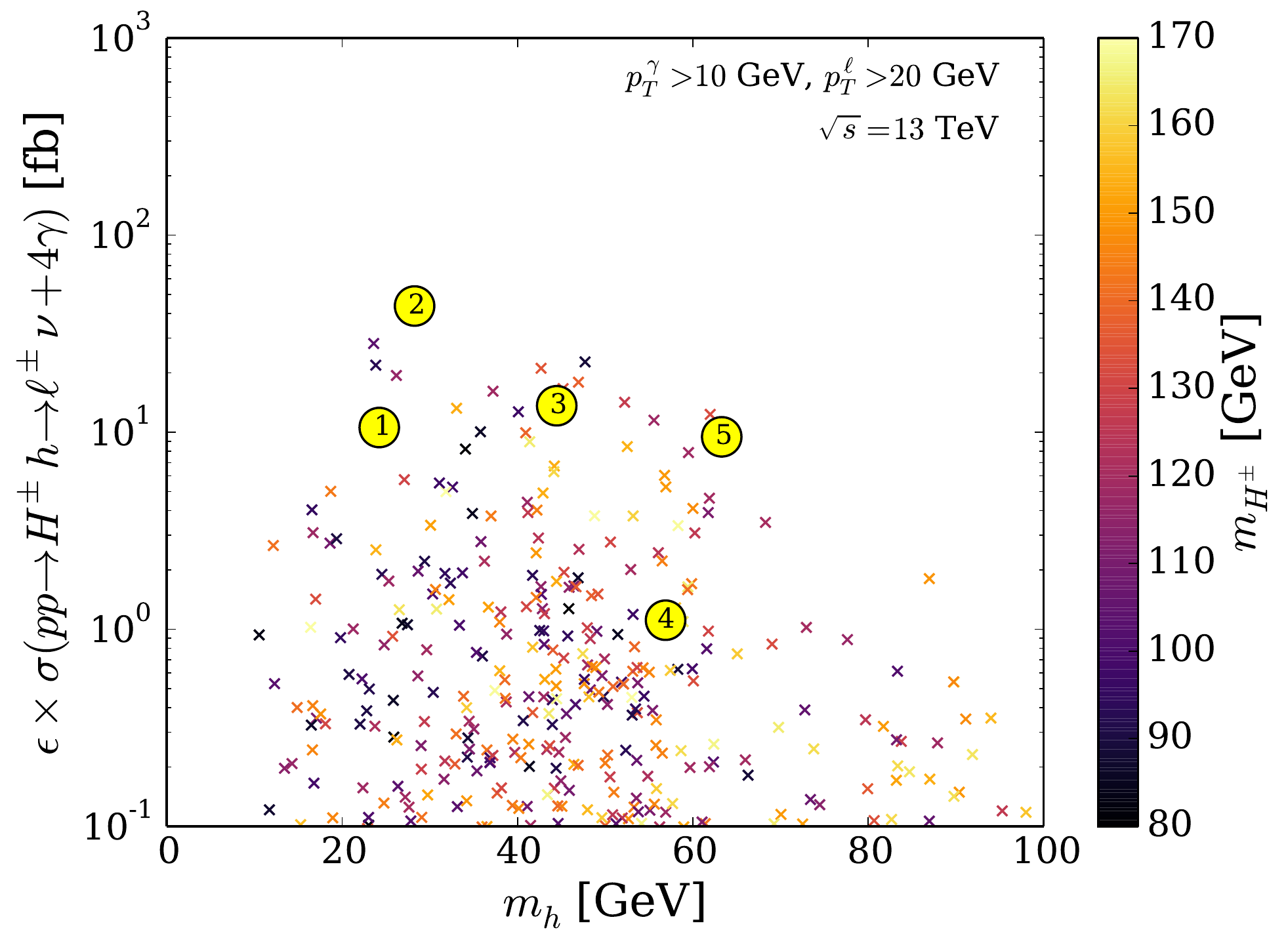}
\includegraphics[width=0.48\textwidth]{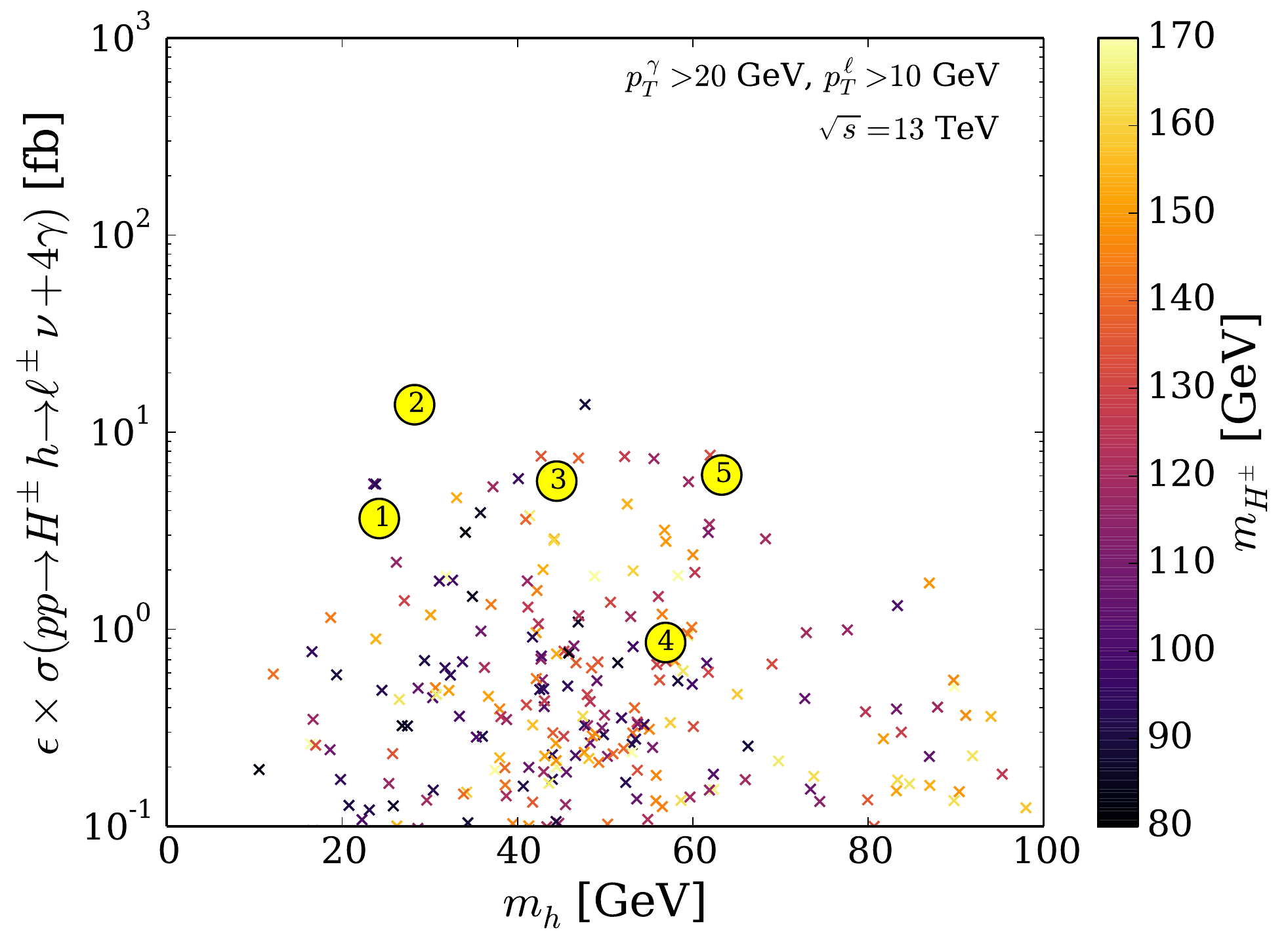}
\caption{Signal cross section $\sigma(\ell^\pm\nu+4\gamma)$ times selection efficiency for the two choices (i) and (ii) of cuts described in the text (left and right, respectively).  The cross section is calculated as $\sigma(\ell^\pm\nu+4\gamma) = \sigma(qq'\to H^\pm h)\times {\rm BR}(H^\pm \to W^\pm h)\times
{\rm BR}(h\to \gamma\gamma)^2\times {\rm BR}(W^\pm\to\ell^\pm\nu)$. The five selected BPs are once again highlighted in yellow circles.}
\label{fig:sigma-w-cuts}
\end{figure}

With this in mind, we consider two sets of cuts: (i) requires that all photons have $p_T^\gamma > 10$\,GeV and the charged lepton has $p_T^\ell > 20$\,GeV, whereas (ii) imposes that $p_T^\gamma > 20$\,GeV and $p_T^\ell > 10$\,GeV. In both cases, we require pseudorapidity $|\eta|<2.5$ for the lepton and each photon, while all objects are required to have an isolation $\Delta R = \sqrt{(\Delta\eta)^2+(\Delta\phi)^2} > 0.4$. To determine the efficiencies of these cuts, we calculated event rates for various masses and determined the corresponding selection efficiencies, $\epsilon=\sigma(\text{cuts})/\sigma(\text{no cuts})$. The results are shown in Fig.~\ref{fig:efficiencies} for both choices of cuts, and demonstrate a strong dependence on the masses involved.
The  effect of these cuts on the signal yield from our scan is shown in Fig.~\ref{fig:sigma-w-cuts}, from which it is clear that, given the negligible background for this process, there is a region of parameter space that should be within reach already at the LHC Run II assuming standard luminosities of order 100\,fb$^{-1}$.

\section{\label{sec:conc} Conclusions} 
In the framework of the 2HDM-I, there exists the possibility of a scenario in which: 
$H$ is the observed CP-even SM-like Higgs boson, 
$h$ is lighter than 125\,GeV (in fact, possibly as light as about 10\,GeV or so), and $H^\pm$ lies in the 80--160 GeV mass range, 
still being consistent with all LHC, LEP Tevatron and $B$-physics data. 
Furthermore, in this scenario, for $\cos\alpha\approx 0$ and some particular choices of the other parameters, $h$ could be highly fermiophobic, decaying fully or dominantly into two photons,  
while the \hpm\ could decay dominantly to $W^\pm h$, escaping the existing
LHC limits on $H^\pm\to$ fermion signatures.

Under these conditions, we have shown that the associated production of the charged Higgs boson with the light
CP-even Higgs, $pp\to H^\pm h$, could be substantial and would lead to a 
$W^\pm + 4 \gamma$ final state with a rather significant event yield. In fact, after reasonable cuts on the $p_T$ of each of the photons and the lepton, under plausible trigger assumptions, alongside those in $\eta$ and $\Delta R$,
the emerging $W^\pm + 4 \gamma$ signal can still enjoy 
a cross section of the order 10--100 fb in an essentially background-free environment.

We therefore look forward to the ATLAS and CMS experiments 
testing this hitherto neglected scenario against their data, as
establishing the signature discussed here will provide not only a direct indication 
of a non-minimal Higgs sector but also circumstantial evidence of a specific 2HDM structure.\\

%%%%% Acknowledgments

\acknowledgments

\noindent The work of RE and SMo is funded through the grant 
H2020-MSCA-RISE-2014 No. 645722 (NonMinimalHiggs). 
SMo is supported in part through the NExT Institute and STFC 
Consolidated Grant ST/J000396/1. SMu acknowledges support from the European Union’s Horizon 2020 grant agreement InvisiblesPlus RISE No.\,690575 for visiting the University of Southampton, where part of this work was carried out. WK thanks Sofia Andringa for helpful feedback. We thank Elin Berge\aa{}s Kuutmann and Richard Brenner for helpful discussions of experimental prospects and Tim Stefaniak for help with HiggsBounds 5.

\appendix
\section{\label{app:halimit} $e^+e^-\to hA$ limit}
The DELPHI collaboration performed a search for the process $e^+e^-\to hA$, with the decays $h\to\gamma\gamma$ and $A\to b\bar{b}$ or $A\to Zh\to Z\gamma\gamma$, when kinematically allowed~\cite{Abdallah:2003xf}. This enabled them to place limits on $\sin^2(\beta-\alpha)$, assuming an exactly fermiophobic $h$, and that $A$ decays entirely into either $b\bar{b}$ or $Zh$. These limits depend on $m_h$ and $m_A$ and are explicitly given for $m_A=50$ GeV and $m_A=115$ GeV. Here, we derive approximate limits for other values of $m_A$, which allows one to constrain models which are not exactly fermiophobic. The centre-of-mass energies, $\sqrt{s}$, and integrated luminosities, $\cal{L}$, used in the analysis are shown in Table~\ref{tab:sl}. For a given model, the expected number of observed events is
\begin{equation}
N_{hA}^\mathrm{exp}(m_h,m_A) = {\rm BR}(h\to\gamma\gamma)\times {\rm BR}(A\to X)\sum_{\{s\}} \sigma_{hA}(s,m_h,m_A){\cal L}(s)\epsilon(s,m_h,m_A).
\end{equation}
Here, $X$ is either $b\bar{b}$ or $Z(h\to\gamma\gamma)$ and $\epsilon(s,m_h,m_A)$ is the signal selection efficiency of the analysis.  If we assume that the variations of the efficiency are not too large, we can replace it with an effective efficiency $\bar{\epsilon}$, which we may then pull outside of the sum and absorb into $\tilde{N}=N_{hA}^\mathrm{exp}/\bar{\epsilon}$, giving

\begin{equation}
\tilde{N}(m_h,m_A) = N_0(m_h,m_A)\cos^2(\beta-\alpha)\times {\rm BR}(h\to\gamma\gamma)\times {\rm BR}(A\to X),
\end{equation}
where
\begin{equation}
N_0(m_h,m_A)=\sum_{\{s\}}\sigma_0(s,m_h,m_A)\times{\cal L}(s).
\end{equation}
Here we have introduced $\sigma_0$, which is the $e^+e^-\to hA$ cross section when $\cos(\beta-\alpha)=0$.

\begin{table}[h]
\begin{center}
\begin{tabular}{|c||r|r|r|r|r|r|r|r|r|}
\hline
$\sqrt{s}$ [GeV] & 182.6 & 188.6 & 191.6 & 195.5 & 199.6 & 201.6 & 205.0 & 206.5 & 206.8 \\\hline
$\cal{L}$ [pb$^{-1}$] & 49.3 & 153.0 & 25.1 & 76.0 & 82.7 & 40.2 & 80.0 & 59.2 & 81.8 \\
\hline
\end{tabular}
\end{center}
\caption{Average centre-of-mass energies and integrated luminosities of the DELPHI analysis.}
\label{tab:sl}
\end{table}

We can then translate a given limit, $s^{\mathrm{lim}}_{\beta\alpha}$, on $\sin(\beta-\alpha)$ into a limit on $\tilde{N}$, given the stated assumptions that BR$(A\to X)=1$ and BR$(h\to\gamma\gamma)$ is for an exactly fermiophobic $h \equiv h_f$:

\begin{equation}
\tilde{N}_{\max}(m_h,m_A) = N_0(m_h,m_A)(1-(s_{\beta\alpha}^{\mathrm{lim}}(m_h,m_A))^2) \times {\rm BR}(h_f\to\gamma\gamma).
\end{equation}

We only have values of $s_{\beta\alpha}^{\mathrm{lim}}$ for $m_h = \{50, 115\}$\,GeV, but since the experimental efficiencies vary slowly~\cite{AndringaDias:2003zz} over our region of interest in the 2HDM-I parameter space, we approximate $\tilde{N}^{\max}$ as a constant. To choose a suitable value, we consider the average values of the two limiting curves over relevant values of $m_h$:

\begin{eqnarray}
\tilde{N}_{\max}^\textrm{avg}(40 < m_h < 90, m_A=50)&=&8.4,\nonumber\\
\tilde{N}_{\max}^\textrm{avg}(25 < m_h < 70, m_A=115)&=&9.1.
\end{eqnarray}

\begin{figure}[h!]
    \begin{center}
    \includegraphics[width=0.48\textwidth]{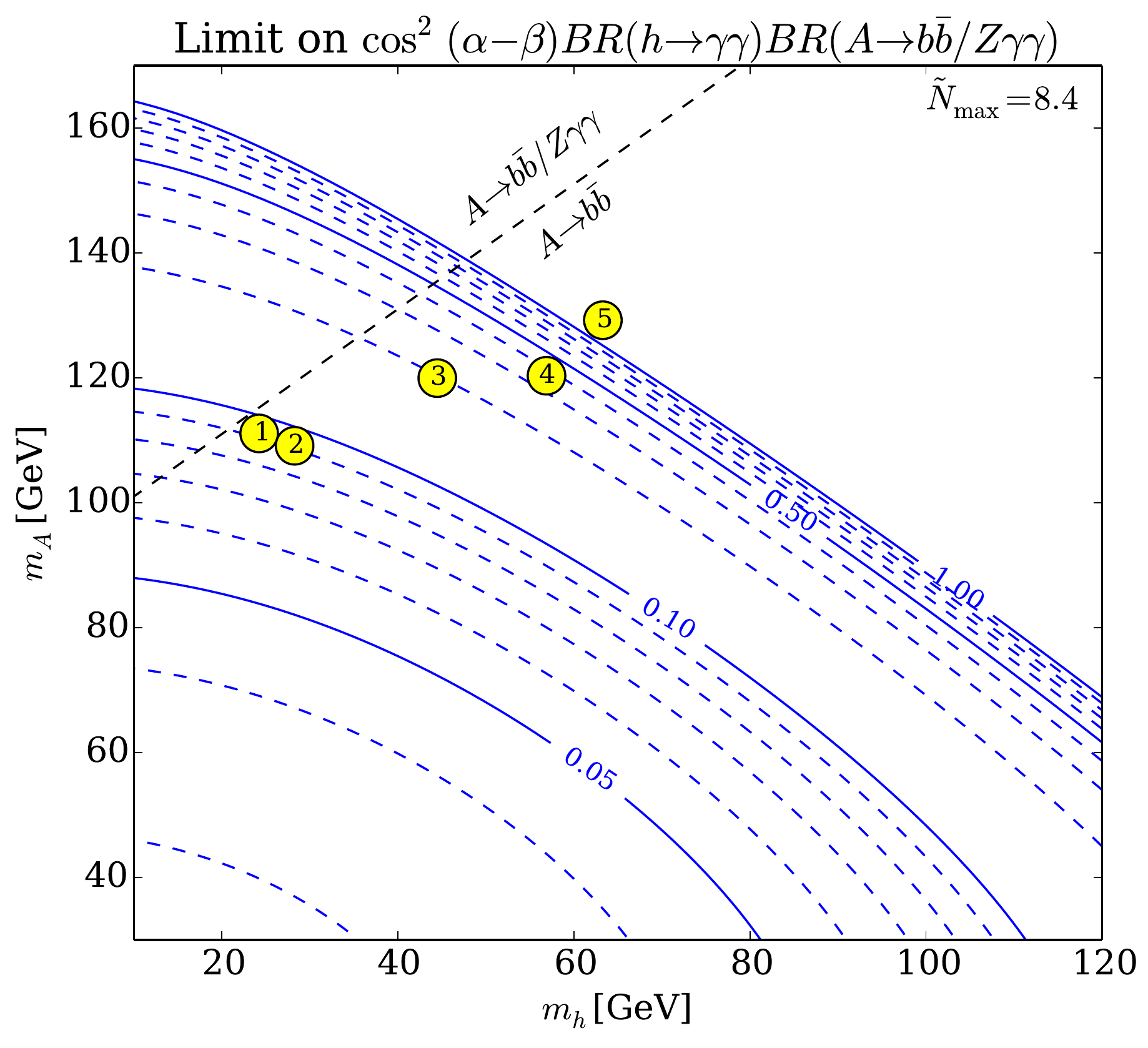}
    \caption{Estimated limits on $\cos^2(\beta-\alpha)\times {\rm BR}(h\to\gamma\gamma)\times {\rm BR}(A\to b\bar{b}/Z\gamma\gamma)$ with $\tilde{N}_{\max}=8.4$. BPs from the text are indicated in yellow circles. The dashed line indicates where $m_A = m_h + m_Z$, above which the on-shell $A\to Z h$ decay is possible.}
    \label{fig:weightedcontours}
    \end{center}
\end{figure}

For $m_A=50$ GeV, we choose a lower limit for $m_h$ of 40\,GeV, as we have only a few points with $m_h + m_A$ much below $m_Z$ due to limits from the $Z$ width measurement. Furthermore, above the upper limit of 90\,GeV very few points have cross sections of interest to this study. For $m_A=115$ GeV, we choose an upper limit of $m_h=70$ GeV, above which $m_h+m_A \gtrsim \sqrt{s}$ for some LEP runs.  The lower limit of 25\,GeV occurs around $m_A = m_h + m_Z$, where the $A\to Z h$ analysis is used. It is notable that both values of $m_A$ give similar results, and for our limit we conservatively choose a value of 8.4.  We may then impose a limit for all values of $(m_h,m_A)$ given by

\begin{equation}
\cos^2(\beta-\alpha) \times {\rm BR}(h\to\gamma\gamma)\times {\rm BR}(A\to X) \leq \frac{\tilde{N}_{\max}}{N_0(m_h,m_A)},
\end{equation}
with $\tilde{N}_{\max}=8.4$.  Finally, we note that the $A\to Z h$ search required an on-shell $Z$ boson, so it is only applicable in the region $m_A > m_h + m_Z$.  In that region, however, the $A\to b\bar{b}$ search is reported to still have sensitivity comparable to the $A \to Z h$ channel, so that, when constraining a particular point in the parameter space, we take the larger of the two:

\begin{equation}
{\rm BR}(A\to X) = 
\begin{cases}
    {\rm BR}(A\to b\bar{b}),& \text{if } m_A < m_h + m_Z\\
    \max{\left\{{\rm BR}(A\to b\bar{b}), {\rm BR}(A\to h Z)\times {\rm BR}(h\to\gamma\gamma)\right \}}  & \text{if } m_A > m_h + m_Z.
\end{cases}
\end{equation}
The resulting limits for $\tilde{N}_{\max}=8.4$ are shown in Fig.~\ref{fig:weightedcontours}.

\section*{References}
\bibliography{W4gamma}
%\bibliographystyle{}

%%%%%%%%%%%%%%%%%%%%%%%%%%%%%%%%%%%%%%%%%%%%%%%%%
\end{document}